\documentclass[aps,prl,twocolumn,groupedaddress]{revtex4}

\usepackage{graphicx}
\usepackage{amsmath}
\usepackage{amssymb}

\def\slantfrac#1#2{ \hspace{3pt}\!^{#1}\!\!\hspace{1pt}/ \hspace{2pt}\!\!_{#2}\!\hspace{3pt}}

\begin{document}

\title{Stable Fractional Vortices in the Cyclic States of Bose-Einstein Condensates}
\author{J.~A.~M.\ Huhtam\"aki,$^{1,2}$ T.~P.~Simula,$^1$ M.~Kobayashi,$^3$ and K.~Machida$^1$}

\affiliation{$^1$Department of Physics, Okayama University, Okayama
700-8530, Japan} \affiliation{$^2$Department of Applied Physics,
Helsinki University of Technology, P.O. Box 5100, 02015 TKK,
Finland} \affiliation{$^3$Department of Physics, University of
Tokyo, Hongo 7-3-11, Bunkyo-ku, Tokyo 113-0033, Japan}

\begin{abstract}
We propose methods to create fractional vortices in the cyclic state
of an $F=2$ spinor Bose-Einstein condensate by manipulating its
internal spin structure using pulsed microwave and laser fields. The
stability of such vortices is studied as a function of the rotation
frequency of the confining harmonic trap both in pancake and cigar
shaped condensates. We find a range of parameters for which the
so-called $\slantfrac{1}{3}$--vortex state is energetically
favorable. Such fractional vortices could be created in condensates
of ${}^{87}{\rm Rb}$ atoms using current experimental techniques
facilitating probing of topological defects with non-Abelian
statistics.
\end{abstract}


\maketitle

The peculiar nature of quantum fluids stands out remarkably when
observing their rotational characteristics. Rigid-body rotation is
forbidden for a superfluid due to the constraint that its velocity
field is bound to be irrotational. Instead they may acquire angular
momentum by hosting a number of quantized vortex
lines~\cite{Donnelly1991}. The superfluid state is characterized in
terms of an order parameter which in general is a multicomponent
complex function. In the single-component (scalar) case, the
quantized vortex lines are phase singularities in the order
parameter field, around which the complex phase of the order
parameter changes by $2\pi \kappa$, where $\kappa \in \mathbb{Z}$ is
the winding number of the vortex.

In systems described by multicomponent order parameters, the vortex
lines may have more complicated structure. A well-known example is
the $A$-phase of ${}^{3}{\rm He}$, in which the Cooper paired
fermionic atoms possess both orbital and spin angular momentum. The
coreless half-integer vortex, for which the phase winding of the
order parameter takes half-integer values, was discovered
theoretically in such system~\cite{Volovik1992}. Existence of these
kinds of topological defects is possible due to the discrete
symmetry under combined gauge transformation and spin rotation of
the related order parameter. Realization of Bose-Einstein
condensates (BECs) in purely optical traps~\cite{Stamper-Kurn1998}
has made it possible to create fractional vortex states also in
ultracold atomic gases. In the absence of strong magnetic fields,
the hyperfine spin degree of freedom of the atoms is unrestricted,
leading to a multicomponent order parameter~\cite{Ohmi1998Ho1998}.

Condensates consisting of optically confined spin-2 atoms are of
special interest here. The first homotopy group $\pi_1$ of the $F=2$
manifold is non-Abelian~\cite{Mermin1979}, enabling non-commutable
composition laws for two vortex lines~\cite{Kobayashi2008}.
Topological defects in the so-called cyclic state have been studied
theoretically in ~\cite{Makela2003Makela2006Semenoff2007}. The
internal symmetry of such order parameter can be mapped to the
discrete tetrahedral group, enabling the existence of fractional
$\slantfrac{1}{3}$--vortices. In this paper, we describe two
plausible schemes to create controllably a
$\slantfrac{1}{3}$--vortex into the cyclic state of an $F=2$ spinor
BEC. Moreover, we investigate the energetic stability of such states
in harmonic traps under rotation, paving the way towards realization
of long-lived non-Abelian topological defects.

An $F=2$ spinor BEC is characterized by an order parameter which has
five spin components, $\psi_k$, $k = {-}2, {-}1, 0, 1, 2$,
corresponding to the eigenstates of the five-dimensional
representation of the spin operator $\hat{S}_z$. The different spin
populations, $N_k = \int d{\bf r} |\psi_k|^2$, can be controlled
accurately in an optical trap in the presence of an offset magnetic
field~\cite{Schmaljohann2004,Kuwamoto2004}.

The ground state phase diagram of an $F=2$ spinor BEC is divided
into ferromagnetic, polar, and cyclic regions depending on the
values of the spin-spin interaction strength $\beta$ and the
spin-singlet coupling constant $\gamma$~\cite{Ciobanu2000}. This is
in close analogy to the phase diagram predicted for $d$-wave paired
Fermi systems~\cite{Mermin1974}. Based on scattering length
measurements, the internal ground state of a ${}^{87}{\rm Rb}$
condensate belongs either to the polar or cyclic
region~\cite{Ciobanu2000,Klausen2001}. Although several studies have
predicted a negative value for $\gamma$, implying that the ground
state would be polar, due to remaining experimental uncertainties it
is not yet clear whether the ground state lies in the cyclic or
polar region~\cite{Widera2006}.

By restricting the atom population into the $\psi_2$ and $\psi_{-1}$
components only, the value of $\gamma$ becomes insignificant: The
spin-singlet pairing interaction does not contribute to the energy
since the spin-singlet pairing amplitude, $\Theta=\sum_k (-1)^k
\psi_k \psi_{-k}$, vanishes. Hence the internal ground state of such
condensate is cyclic.
In particular, let us consider states of the form~\cite{comment1}
\begin{eqnarray}
\begin{pmatrix} \psi_{+2}(r,\varphi,z) \\ \psi_{-1}(r,\varphi,z) \end{pmatrix}
= e^{i \vartheta} R_z(\phi) \begin{pmatrix} \psi_{+2}(r,z) \\
\psi_{-1}(r,z) \end{pmatrix},
\end{eqnarray}
where $(r,\varphi,z)$ denote the cylindrical coordinates, and
$\vartheta$ and $\phi$ are functions of the azimuthal angle
$\varphi$. $R_z(\phi)=\textrm{diag}(\exp[2i\phi], \exp[-i\phi])$
describes a spin rotation through an angle $\phi$ about the $z$-axis
and $e^{i \vartheta}$ is a gauge transformation. By choosing
$\vartheta=\phi=\varphi/3$, we obtain a state with unit~(zero) phase
winding in the $\psi_{+2}~(\psi_{-1})$ component. For such state, by
traversing along a loop encircling the origin, the order parameter
is continuously transformed onto itself accompanied by a spin
rotation of one-third of a full circle and a phase change of
$\frac{2\pi}{3}$. This configuration is therefore called a
$\slantfrac{1}{3}$--vortex state. Similarly, choosing
$\vartheta=2\varphi/3$, $\phi=-\varphi/3$, yields a
$\slantfrac{2}{3}$--vortex state.

Here we restrict to mean-field theory for which the zero-temperature
energy of the condensate in the absence of external magnetic fields
is given by~\cite{Ciobanu2000}
\begin{eqnarray}
\label{energy} E = \int d{\bf r} \left[ \psi^*_k \hat{h}_\Omega
\psi_k + \frac{\alpha}{2} n^2 + \frac{\beta}{2} \langle {\bf f}
\rangle^2 + \frac{\gamma}{2} |\Theta|^2 \right],
\end{eqnarray}
with implied summation over repeated indices. In Eq.~(\ref{energy}),
$\hat{h}_{\Omega} = -\frac{\hbar^2 }{2m} \nabla^2 + V_{\rm
trap}({\bf r})- {\bf \Omega} \cdot \hat{{\bf L}}$ is the
single-particle Hamiltonian where $m$ is the mass of a ${}^{87}{\rm
Rb}$ atom, $V_{\rm trap}$ the external optical potential, $\hat{{\bf
L}}$ the angular momentum operator, and $\Omega$ the rotation
frequency of the trap. The total density of the condensate is
denoted by $n = \psi_k^* \psi_k$, and the spin-spin interaction
energy by $\langle {\bf f} \rangle^2 = \psi_i^* \psi_k^*
f^\delta_{ij} f^\delta_{kl} \psi_j \psi_l$, where $f^\delta_{kl}$ is
a matrix element of $\hat{S}_\delta/\hbar$. The parameters $\alpha$,
$\beta$, and $\gamma$ are related to the $s$-wave scattering lengths
$a_F$ in the total hyperfine state $F$ according to $\alpha=
\frac{1}{7}(4g_2 + 3g_4)$, $\beta=-\frac{1}{7}(g_2 - g_4)$, and
$\gamma=\frac{1}{5}(g_0 - g_4) - \frac{2}{7}(g_2-g_4)$, where
$g_F=\frac{4 \pi \hbar^2}{m} a_F$ are the bare coupling
constants~\cite{Ciobanu2000,Koashi2000}.

Minimizing the energy functional, Eq.~(\ref{energy}), with a fixed
total number of particles, leads to the zero-temperature
Gross-Pitaevskii equation for the order parameter
\begin{eqnarray}
\label{GP} \hat{h}_\Omega \psi_k + \alpha n \psi_k + \beta \langle
f^\delta \rangle f^\delta_{kl} \psi_l + \gamma \Theta (-1)^k
\psi_{-k}^* = \mu \psi_k,
\end{eqnarray}
where $\mu$ is the chemical potential of the system. We have solved
Eq.~(\ref{GP}), both in pancake and cigar-shaped harmonic traps, in
order to find the ground states of the system. In the former case,
we choose the axial trapping frequency $\omega_z$ large enough so
that the condensate dynamics is frozen in the flattened direction,
and hence the axial profile of the wave function can be accurately
approximated by a Gaussian function. In this limit, the interaction
parameter $\alpha$ is replaced by the quasi-2D form $\alpha_{\rm 2D}
= 2\sqrt{2\pi} \hbar \omega_z a_r \left[ \frac{1}{7} (4 a_2 + 3 a_4)
\right]$. Setting the radial and axial trapping frequencies to
$(\omega_r/2\pi,\omega_z/2\pi)=(100,1000)\, {\rm Hz}$, respectively,
the harmonic oscillator length in the radial direction is $a_r=1.1\,
\mu {\rm m}$, and we have chosen the value of $\alpha_{\rm 2D}$ to
correspond to $2 \times 10^4$ ${}^{87}{\rm Rb}$ atoms using the
scattering length presented in~\cite{Ciobanu2000}. Moreover, we set
$\beta = \alpha/100,$ $(\gamma = -\alpha/187)$ according to the
values measured in~\cite{Widera2006}.

\begin{figure*}[!t]
\includegraphics[width=500pt]{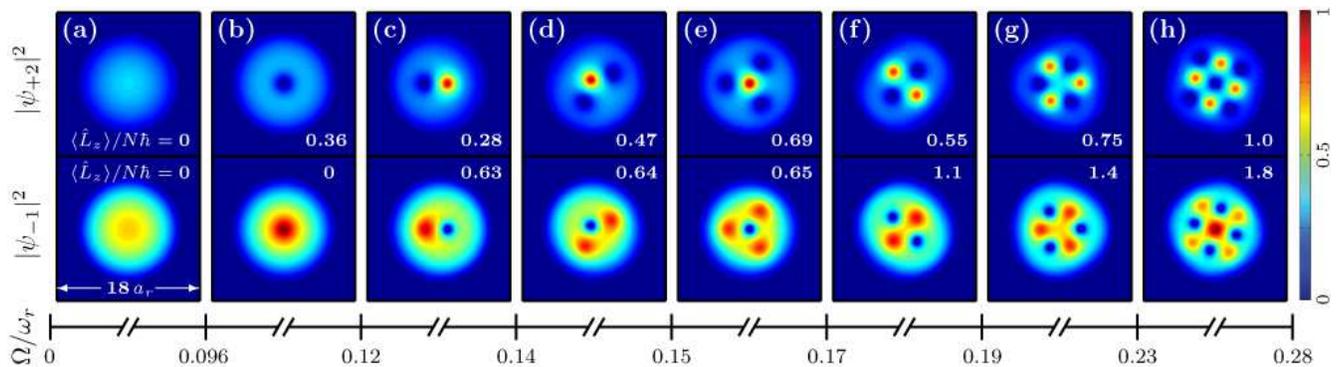}
\caption{(Color online) Ground state density profiles of
${}^{87}{\rm Rb}$ atoms in the $|F=2,m_F=+2 \rangle$ component
(upper row) and $|F=2,m_F=-1 \rangle$ component (lower row) in a
harmonic trap rotating at frequency $\Omega$ under the constraint
that only these components are populated. The density minima in the
$|{+}2\rangle$ component correspond to fractional
$\slantfrac{1}{3}$--vortices, and those in $|{-}1\rangle$ to
$\slantfrac{2}{3}$--vortices. The $\slantfrac{1}{3}$--vortex is the
ground state within the interval $\Omega \in [0.0957\, \omega_r,
0.123\, \omega_r]$ for the chosen parameter values. The color bar is
scaled with respect to the maximum of the total density, ${\rm max}
\left[n({\bf r})\right]$.} \label{gs}
\end{figure*}

Figure~\ref{gs} depicts the computed ground state profiles of the
condensate for a range of rotation frequencies, $\Omega$, under the
constraint that only the $|{+}2\rangle$ and $|{-}1\rangle$
components have non-zero population. The top and bottom rows show
the density of particles in the $|{+}2\rangle$ and $|{-}1\rangle$
components, respectively. The field of view is $20\, a_r \times 20\,
a_r$, and the angular momentum of each component is given inside the
figure. The horizontal bar below the figure shows the critical
rotation frequencies for which the energies of neighboring ground
state configurations are equal. The ground states are qualitatively
similar for different values of $\Omega$ within each stability
window, and the representative profiles shown refer to the rotation
frequency at the middle of the interval. However, the total angular
momentum and magnetization $\langle f_z \rangle$ of the condensate
varies within each stability window because the population ratio
depends on the rotation frequency. For example, for the
$\slantfrac{1}{3}$--vortex state, illustrated in Fig.~1(b), the
angular momentum ranges from $0.34\, N\hbar$ to $0.40\, N\hbar$ and
the magnetization from
$0.031\,N$ to $0.099\,N$.

For the chosen parameters, the $\slantfrac{1}{3}$--vortex state
$\left[\psi_{+2}(r)e^{i\varphi}, \psi_{-1}(r)\right]$ turns out to
be energetically favorable compared to the non-rotating ground state
when $\Omega > 0.0957\, \omega_r$. As the rotation frequency is
further increased above $\Omega=0.123\, \omega_r$, a pair of
$\slantfrac{1}{3}$-- and $\slantfrac{2}{3}$--vortices occupies the
ground state. This state is analogous to the vortex molecule
structure discussed
in~\cite{Kasamatsu2004} within the nonlinear sigma model. 
In the present context, the $\left(
\slantfrac{1}{3},\slantfrac{2}{3} \right)$--vortex pair can be
viewed as a split gauge vortex, because the total spin rotation
along a loop containing the pair vanishes and the gauge fields of
the individual vortices add up to an integer phase winding~$e^{i\varphi}$.
The $\slantfrac{2}{3}$--vortex state carries roughly twice the
angular momentum and kinetic energy of the
$\slantfrac{1}{3}$--vortex state. The former becomes energetically
favorable compared to the non-rotating ground state at
$\Omega=0.111\, \omega_r$, and at $\Omega=0.127\, \omega_r$ when
compared to the $\slantfrac{1}{3}$--vortex state. However, this
state is not the ground state of the system for any value of
$\Omega$. For comparison, in the scalar $|F=2,m_F={+}2 \rangle$
condensate with the same parameter values, the singly quantized
vortex state is the ground state within the interval $\Omega \in
[0.164\, \omega_r, 0.222\, \omega_r]$. We have also confirmed that
other possible vortex states in the given configuration, such as
gauge or integer spin vortices, are energetically unstable.

At even higher rotation frequencies additional $\slantfrac{1}{3}$---
and $\slantfrac{2}{3}$--vortices enter the system, yielding a rich
structure in the ground state phase diagram, see
Figs.~\ref{gs}(d)--(h). With fast enough rotation, lattices
consisting of $\slantfrac{1}{3}$-- and $\slantfrac{2}{3}$--vortices
appear. Vortex lattice transitions in such setup have been studied
recently~\cite{Barnett2008}. Moreover, vortex lattice structures in
two-component condensates have previously been studied
theoretically~\cite{Mueller2002,Kasamatsu2003}, as well as observed
experimentally~\cite{Schweikhard2004}. In the present study, the
population ratio is $N_{+2}/N_{-1} \sim 0.5$, instead of being close
to unity, which is inherited from the underlying magnetic state
(cyclic). For the parameter values of ${}^{87}{\rm Rb}$, we observe
only interlaced square and rectangular vortex lattices. The shape of
the unit cell depends on the rotation frequency as well as on the
total particle number $N$, tending more towards a square lattice in
the limit $\Omega \to \omega_r$, and $N \to \infty$. For example,
with $\Omega = 0.90\, \omega_r$ and $N=2 \times 10^4$, both unit
cells are present: a square lattice close to the trap center where
particle density is high, bounded by rectangular unit cells closer
to the periphery of the cloud. The magnetization of the interlaced
square lattice resembles notably that of the ground state of a 2D
Ising model with antiferromagnetic interactions. Moreover, these
vortex lattices can be viewed as consisting of $\left(
\slantfrac{1}{3}, \slantfrac{2}{3} \right)$--vortex pairs, such as
depicted in Fig.~\ref{gs}(c).

In the cigar-shaped limit, we choose the trapping frequencies
$(\omega_r/2\pi,\omega_z/2\pi)=(100,10)\, {\rm Hz}$, and the
coupling constants $\alpha, \beta, (\gamma)$ corresponding to $3
\times 10^5$ ${}^{87} {\rm Rb}$ atoms.
The $\slantfrac{1}{3}$--vortex state is the ground state within the
interval $\Omega \in [0.171\, \omega_r, 0.237\, \omega_r]$.
The vortex line is bent in the stationary state of the system even
in cylindrically symmetric confinement, analogously to the scalar
case~\cite{Garcia-Ripoll2001}. Hence, the angular momentum of the
$\slantfrac{1}{3}$--vortex state is not determined only through the
populations in the two components, $N_{+2}$ and $N_{-1}$, but is
strongly affected also by the bending of the vortex line. Within the
stability window, the atom number in the $|{+}2\rangle$ component
ranges from $0.33\,N$ to $0.41\,N$, whereas the total angular
momentum ranges from $0.21\, N\hbar$ to $0.39\, N\hbar$. For
comparison, the singly quantized vortex state in the scalar
$|F=2,m_F={+}2 \rangle$ case is energetically favorable within the
interval $\Omega \in [0.294\, \omega_r, 0.394\, \omega_r]$.

\begin{figure}[!t]
\includegraphics[width=242pt]{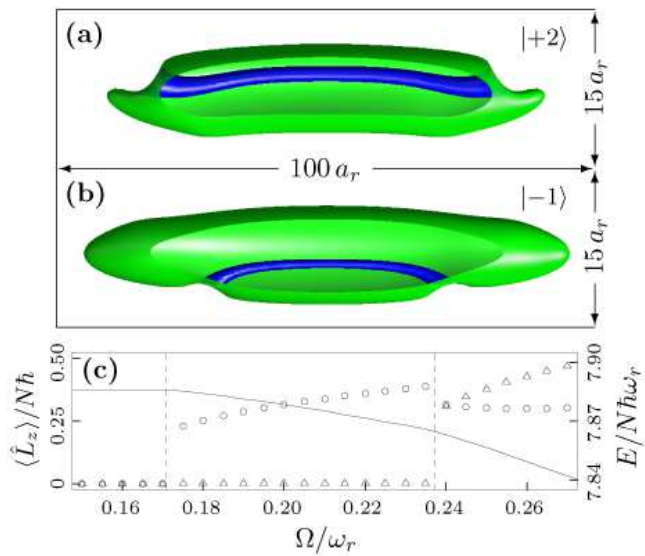}
\caption{(Color online) {\bf (a)} Isosurface plot of the density of
particles in the $|{+}2\rangle$ component $|\psi_{+2}({\bf r})|^2$,
and {\bf (b)} in the $|{-}1\rangle$ component $|\psi_{-1}({\bf
r})|^2$. The field of view is illustrated approximately by the
double arrows. In both isosurfaces, the density is equal to $10\%$
of the maximum total density of particles, ${\rm max} \left[n({\bf
r})\right]$. For the chosen rotation frequency of
$\Omega=0.25\,\omega_r$, the ground state contains a
$\left(\slantfrac{1}{3}, \slantfrac{2}{3} \right)$--vortex pair. The
vortex lines, for clarity drawn with different color as the rest of
the isosurface, are bent even in an axisymmetric trap. {\bf (c)}
Angular momentum in the $|{+}2\rangle$ component (circles) and in
the $|{-}1\rangle$ component (triangles) as a function of the
rotation frequency $\Omega$. The solid curve shows the ground state
energy of the condensate. The $\slantfrac{1}{3}$--vortex is the
ground state configuration between the dashed vertical lines.}
\label{3D}
\end{figure}

As the rotation frequency is increased further, the ground state
contains a $\left( \slantfrac{1}{3}, \slantfrac{2}{3}
\right)$--vortex pair which is illustrated in the isosurface plots
of Fig.~\ref{3D} for $\Omega=0.25\, \omega_r$. At each point on the
surface shown in Fig.~\ref{3D}(a), the particle density in the
$|{+}2\rangle$ component is equal to $10\%$ of the total particle
density, and similarly for the $|{-}1\rangle$ component in
Fig.~\ref{3D}(b). The surfaces are cut open to clarify the vortex
lines inside the clouds.

The ground state energy as a function of rotation frequency $\Omega$
is plotted in Fig.~\ref{3D}(c) (the solid curve). The interval
within which the $\slantfrac{1}{3}$--vortex is stable is bounded by
the dashed vertical lines. The slope of the energy curve steepens as
the total angular momentum of the state increases. The angular
momentum of each component is shown with circles for $|{+}2\rangle$
and triangles for $|{-}1\rangle$. Angular momentum of the
$|{-}1\rangle$ component deviates slightly from zero even in the
$\slantfrac{1}{3}$--vortex state since bending of the vortex line
breaks the cylindrical symmetry in both components. For higher
rotation frequencies, a square lattice of vortices forms also in the
cigar-shaped case.

Finally, we will describe two alternative methods of creating a
$\slantfrac{1}{3}$--vortex state in a trapped condensate. The first
scheme is a direct generalization of the method utilized by Matthews
{\it et al.} to create the first vortex in a two-component
BEC~\cite{Williams1999,Matthews1999}.
The condensate is first prepared in the $|F=1,m_F=0\rangle$ state in
an optical trap, e.g., by applying a magnetic field gradient during
the evaporative cooling stage~\cite{Chang2004}. Subsequently,
two-photon microwave and auxiliary laser fields are used, as in
\cite{Matthews1999}, to transfer roughly one-third of the population
directly into a finite orbital angular momentum state in
$|F=2,m_F={+}2\rangle$. Thereafter, the desired configuration in the
$F=2$ manifold is achieved by switching off the laser beam and
transferring the remaining population from $|F=1,m_F=0\rangle$ into
$|F=2,m_F=-1\rangle$ by a resonant microwave field. The scheme
described above yields a $\slantfrac{1}{3}$--vortex state
$[\psi_{+2}(r,z)e^{i\varphi},\psi_{-1}(r,z)]$ with the desired
population ratio of $N_{+2}/N_{-1} \sim 0.5$. Following the same
prescription but initially preparing the atoms in
$|F=1,m_F={-}1\rangle$ results in
$[\psi_{+1}(r,z)e^{i\varphi},\psi_{-2}(r,z)]$ which is a
$\slantfrac{2}{3}$--vortex state. However, in this case some of the
initially prepared atoms would have to be discarded in order to
obtain the optimal population ratio with maximal transfer of $50\%$
with the two-photon microwave field. The auxiliary far-detuned
rotating laser beam could in principle be utilized in rotating the
condensate~\cite{Madison2000}, providing stability for the
fractional vortex state.

Another possibility to create the $\slantfrac{1}{3}$--vortex state
employs a Laguerre-Gaussian (LG) laser mode which has previously
been utilized for creating multiply quantized vortices in a scalar
BEC as well as vortex-antivortex superposition states in spinor
condensates~\cite{Andersen2006,Wright2009}.
The $\slantfrac{1}{3}$--vortex configuration could be achieved by
preparing the condensate initially in $|F=1,m_F=0\rangle$, and
transferring roughly one-third of the atoms into
$|F=2,m_F=+2\rangle$ using counterpropagating $\sigma^+$ polarized
LG and $\sigma^-$ polarized Gaussian laser beams
~\cite{Wright2009}. The remaining population in
$|F=1,m_F=0\rangle$ could be transferred into $|F=2,m_F={-}1\rangle$
using resonant microwave pulses yielding the desired vortex state
and population ratio.

In experiments, the lifetime of a condensate in the unstable $F=2$
manifold has been limited to a few hundred milliseconds due to
inelastic collisions~\cite{Schmaljohann2004}. Therefore, it seems
more promising to create the $\slantfrac{1}{3}$--vortex dynamically
instead of relying on the cloud to thermalize into the lowest energy
state. The magnetization relaxation time scale is beyond current
lifetimes of the condensates, and hence the population ratio of the
$|{+}2\rangle$ and $|{-}1\rangle$ components may be expected to be
conserved even in the presence of weak magnetic stray fields. Moreover, very slow spin-mixing dynamics has been
reported in ${}^{87}{\rm Rb}$ condensates initially prepared in the
cyclic configuration~\cite{Schmaljohann2004}, in favor of stable fractional vortex states.

In summary, we have proposed two realistic methods to create a
fractional $\slantfrac{1}{3}$--vortex state in an $F=2$ spinor
Bose-Einstein condensate in a laboratory experiment. Both schemes
are based on transferring population from the stable $F=1$ manifold
into two spin states, $|m_F={+}2\rangle$ and $|m_F={-}1\rangle$, in
the $F=2$ manifold by using a combination of laser beams and pulsed
magnetic fields. We have performed energetic stability analysis of
such fractional vortex states, and found that, e.g., the
$\slantfrac{1}{3}$--vortex is the energetically favored state within
an interval of rotation frequencies of the confining potential. For
increasing rotation frequencies, the ground state contains more and
more of $\slantfrac{1}{3}$-- and $\slantfrac{2}{3}$--vortices, which
eventually form an interlaced square lattice. The energetic
stability of the fractional vortices under rotation raises hope for
experimental realization of topological defects with non-Abelian
statistics. Effects related to the non-Abelian nature of the
fractional vortices could be probed for instance by vortex-vortex
collision experiments~\cite{Kobayashi2008}.

T.~Hirano, T.~Mizushima and M.~Takahashi are acknowledged for useful
discussions. The work was financially supported by JSPS and the
V\"ais\"al\"a Foundation.


\begin{thebibliography}{99}

\bibitem{Donnelly1991} R.~J.~Donnelly, {\it Quantized Vortices in Helium
II}, (Cambridge University Press, Cambridge, 1991).


\bibitem{Volovik1992} G.~E.~Volovik, {\it Exotic properties of superfluid ${}^{3}{\rm
He}$}, (World Scientific, Singapore, 1992).

\bibitem{Stamper-Kurn1998} D.~M.~Stamper-Kurn {\it et al.}, Phys.
Rev. Lett. {\bf 80}, 2027 (1998).

\bibitem{Ohmi1998Ho1998} T.~Ohmi and K.~Machida, J.~Phys.~Soc.~Jpn. {\bf 67}, 1822
(1998); T.-L.~Ho, Phys. Rev. Lett. {\bf 81}, 742 (1998).


\bibitem{Mermin1979} N.~D.~Mermin, Rev. Mod. Phys. {\bf 51}, 591
(1979).

\bibitem{Kobayashi2008} M.~Kobayashi {\it et al.},
Phys.~Rev.~Lett. {\bf 103}, 115301 (2009).

\bibitem{Makela2003Makela2006Semenoff2007} H.~M\"akel\"a {\it et al.},
J.~Phys.~A {\bf 36}, 8555 (2003); H.~M\"akel\"a, J.~Phys.~A {\bf
39}, 7423 (2006); G.~W.~Semenoff and F.~Zhou, Phys. Rev. Lett. {\bf
98}, 100401 (2007).



\bibitem{Schmaljohann2004} H.~Schmaljohann {\it et al.}, Phys. Rev. Lett. {\bf 92}, 040402 (2004).

\bibitem{Kuwamoto2004} T.~Kuwamoto {\it et al.}, Phys. Rev. A {\bf
69}, 063604 (2004).


\bibitem{Ciobanu2000} C.~V.~Ciobanu {\it et al.}, Phys. Rev.
A {\bf 61}, 033607 (2000).

\bibitem{Mermin1974} M.~D.~Mermin, Phys. Rev. A {\bf 9}, 868 (1974).

\bibitem{Klausen2001} N.~N.~Klausen {\it et al.},
Phys. Rev. A, {\bf 64}, 053602 (2001).

\bibitem{Widera2006} A.~Widera {\it et al.}, New J. Phys. {\bf 8}, 152 (2006).

\bibitem{comment1} Without loss of generality, we omit writing the relative phase between the
components and restrict the discussion to involve only
$|{+}2\rangle$ and $|{-}1\rangle$ states whereas all of our results
apply also to the $|{-}2\rangle$ and $|{+}1\rangle$ combination.

\bibitem{Koashi2000} M.~Koashi and M.~Ueda, Phys. Rev. Lett. {\bf
84}, 1066 (2000).


\bibitem{Kasamatsu2004} K.~Kasamatsu {\it et al.}, Phys.
Rev. Lett. {\bf 93}, 250406 (2004).



\bibitem{Barnett2008} R.~Barnett {\it et al.},
Phys. Rev. Lett. {\bf 100}, 240405 (2008).

\bibitem{Mueller2002} E.~J.~Mueller and T.-L.~Ho, Phys. Rev. Lett.
{\bf 88}, 180403 (2002).

\bibitem{Kasamatsu2003} K.~Kasamatsu {\it et al.}, Phys.
Rev. Lett. {\bf 91}, 150406 (2003).

\bibitem{Schweikhard2004} V.~Schweikhard {\it et al.}, Phys. Rev. Lett. {\bf 93}, 210403 (2004).

\bibitem{Garcia-Ripoll2001} J.~J.~Garc\'ia-Ripoll and
V.~M.~P\'erez-Garc\'ia, Phys. Rev. A {\bf 64}, 053611 (2001).

\bibitem{Williams1999} J.~E.~Williams and M.~J.~Holland, Nature (London) {\bf
40}, 568 (1999).

\bibitem{Matthews1999} M.~R.~Matthews {\it et al.}, Phys. Rev. Lett. {\bf
83}, 2498 (1999).

\bibitem{Chang2004} M.-S.~Chang {\it et al.}, Phys. Rev. Lett. {\bf 92}, 140403 (2004).

\bibitem{Madison2000} K.~W.~Madison {\it et al.}, Phys. Rev. Lett. {\bf 84}, 806 (2000).

\bibitem{Andersen2006} M.~F.~Andersen {\it et al.}, Phys. Rev. Lett. {\bf 97}, 170406 (2006).

\bibitem{Wright2009} K.~C.~Wright {\it et al.}, Phys. Rev. Lett. {\bf 102}, 030405 (2009).


\end{thebibliography}
\end{document}